\documentclass[twocolumn, aps]{revtex4-2}
\usepackage{graphicx}
\usepackage{amsmath}
\usepackage{graphics}
\usepackage{epsfig}

\usepackage{amsmath}
\usepackage{amsfonts}
\usepackage{amssymb}
\usepackage{xcolor}

\usepackage[normalem]{ulem}

\begin{document}

\newcommand{\lyxdot}{.}

\title{Non-equilibrium lifetimes of DNA under electronic current in a molecular junction}

\author{Julian A. Lawn}
 \affiliation{College of Science and Engineering,  James Cook University, Townsville, QLD, 4811, Australia }
 
 \author{Nicholas S. Davis}
 \affiliation{College of Science and Engineering,  James Cook University, Townsville, QLD, 4811, Australia }
 
\author{Daniel S. Kosov}
 \email{daniel.kosov@jcu.edu.au}
 \affiliation{College of Science and Engineering,  James Cook University, Townsville, QLD, 4811, Australia }

\begin{abstract}
We investigate the non-equilibrium mechanical motion of double-stranded DNA in a molecular junction under electronic current using Keldysh-Langevin molecular dynamics.  Non-equilibrium electronic force reshapes the effective potential energy surface, and along with electronic viscosity force and stochastic force, governs  voltage-dependent dynamics of DNA's collective mechanical coordinate. We compute mean first-passage times to quantify the non-equilibrium lifetime of the DNA junction. At low voltage biases, electron-mechanical motion coupling destabilises DNA by shifting the potential minimum towards critical displacement and suppressing barriers, shortening lifetimes by several orders of magnitude. Unexpectedly, at higher voltages the trend reverses: the potential minimum shifts away from instability and the barrier re-emerges, producing re-stabilisation of the junction. In addition, we demonstrate the Landauer blowtorch effect in this system: coordinate-dependent fluctuations generate a spatially varying effective temperature, changing current-induced dynamics of mechanical degrees of freedom. Apparent temperatures of DNA mechanical motion increase far above ambient due to current-induced heating, correlating with suppressed electronic current at stronger couplings. Our results reveal a non-equilibrium interplay between current-driven forces, dissipation, and fluctuations  in DNA junctions, establishing mechanisms for both destabilisation and recovery of DNA stability under electronic current.

\end{abstract}
 
\maketitle

\section{Introduction}

The mechanical stability of DNA is central to its biological role as the carrier of genetic information. While the double helix is stabilised by hydrogen bonding and base stacking, controlled local instabilities enable replication, transcription, and repair. Beyond biology, DNA has become a versatile component in nanotechnology -  serving as a highly programmable scaffold for building nanoscale materials (e.g. structural assemblies and devices), and as a core element in molecular electronic applications such as single-molecule conductors and junctions \cite{moletronics}. In such junctions, DNA is not only exposed to environmental fluctuations but also to tunnelling electronic currents, raising fundamental questions if current flow can be used to control DNA structural stability and instability.

It is well established that electric fields and currents can influence the  structure of the molecule in an electronic junction, from field-induced chemical reactions to current-driven conformational switching \cite{harzmann2015,darwish2016,schwarz2016,reimers2023}. We propose the microscopic mechanisms by which tunnelling electrons destabilise -- or in some cases stabilise -- DNA in a molecular junction environment.  Quantum electronic current couples to classical mechanical motion of the DNA.  Such coupling injects significant energy into mechanical motion, modifies the potential energy surface, and generates spatially varying fluctuations and dissipation, leading to complex and sometimes counterintuitive effects.

In this work, we use Keldysh-Langevin molecular dynamics \cite{kershaw2020,preston2020,preston2023}  to quantify the stability, instability and non-equilibrium lifetime of DNA molecular junctions under voltage bias. We focus on the dynamics of a collective mechanical coordinate and measure stability using mean first-passage times to a critical displacement threshold. This allows us to connect microscopic current-induced forces to macroscopic lifetimes of the junction. 
Since DNA is a chiral molecule, we are also interested in how its mechanical motion with non-equilibrium current induced forces affects the spin-resolved electronic current. 
Our results reveal three central findings: (i) tunnelling electrons can dramatically reduce DNA lifetimes by lowering barriers and shifting potential minima; (ii) at higher voltages, stability can paradoxically recover as barriers re-emerge; and (iii) coordinate-dependent noise gives rise to 
a Landauer blowtorch effect \cite{landauer1975,landauer1993}, where locally enhanced fluctuations dramatically change rates of chemical reactions.

These results provide a new perspective on current-driven dynamics in DNA and related biomolecular junctions. More broadly, they illustrate how tunnelling electrons reshape molecular stability through an interplay of forces, dissipation, and noise -- non-equilibrium mechanisms that are likely to be relevant across a broad range of molecular electronic junctions.

The paper is organised as follows. In Section II, we present the theoretical framework, including the DNA model and the Keldysh-Langevin molecular dynamics.  Section III presents the results: voltage-dependent effective potentials, non-equilibrium lifetimes, and the emergence of the Landauer blowtorch effect.  Finally, Section IV concludes with a summary of the main findings.

\section{Computational model}

\subsection{Hamiltonian}

We adopt the Hamiltonian which we used in our previous paper on DNA's mechanical motion \cite{davis2024}. We consider a system composed of a double-stranded DNA molecule connected at both ends to macroscopic electrodes, forming a molecular junction. Each nucleotide within the DNA is modelled as a single electronic spin-orbital, and electrons are allowed to tunnel both along individual strands (intra-strand hopping) and across complementary bases within a base pair (inter-strand hopping). The electronic properties of the molecule are conformation-dependent, with the nuclear dynamics encapsulated by a single classical mechanical degree of freedom.

 The total Hamiltonian of the system is expressed as
\begin{equation}
H = H_{M}(p,x) + H_{L} + H_{R} + H_{ML} + H_{MR},
\end{equation}
where 
$H_{L}$ and $H_{R}$ represent the left and right electrodes, respectively, and $H_{ML}$ and $H_{MR}$ describe the electronic coupling between the DNA and electrodes. The DNA Hamiltonian
\begin{equation}
H_{M} = H_{\text{el}} + H_{\text{e-mech}}(x) + H_{\text{mech}}(p, x),
\end{equation}
consists of electronic Hamiltonian $H_{\text{el}}$,
the term $H_{\text{e-mech}}(x)$ accounts for the coupling between electronic degrees of freedom and the mechanical motion of DNA, while $H_{\text{mech}}(p, x)$ describes the dynamics of the mechanical coordinate itself. 

The electrodes are modelled as non-interacting electron reservoirs:
\begin{equation}
H_{L} + H_{R} = \sum_{\alpha k \sigma} \epsilon_{\alpha k \sigma} d_{\alpha k \sigma}^{\dagger} d_{\alpha k \sigma},
\end{equation}
where $d_{\alpha k \sigma}^{\dagger}$ ($d_{\alpha k \sigma}$) creates (annihilates) an electron in state $k$ with spin $\sigma$ in electrode $\alpha \in {L, R}$.

The electronic Hamiltonian of the DNA molecule incorporates intra-strand and inter-strand hopping as well as spin-orbit coupling (SOC), following the model of Refs. \cite{guo2012, du2020}:
\begin{align}
H_{M} &= \sum_{\beta j \sigma} \epsilon_{\beta} d_{\beta j \sigma}^{\dagger} d_{\beta j \sigma} +
\sum_{\beta j \sigma} t_{\beta} \left( d_{\beta j \sigma}^{\dagger} d_{\beta, j+1, \sigma} + \text{h.c.} \right) \nonumber \\
&\quad + \sum_{j \sigma} v \left( d_{A j \sigma}^{\dagger} d_{B j \sigma} + \text{h.c.} \right) + V_{\text{SOC}}.
\end{align}
Here, $j$ indexes base pairs along the chain, while $\beta \in {A, B}$ labels the DNA strands. The on-site energies are denoted by $\epsilon_{\beta}$, with intra-strand hopping integrals $t_{\beta}$ and inter-strand hopping $v$. Molecular parameters are the same as in our previous paper \cite{davis2024}: $\epsilon_A = -0.2$ eV, $\epsilon_B = 0.1$ eV, $t_A = 0.1$ eV, $t_B = -0.14$ eV, and $v = -0.08$ eV.{  The number of DNA basepairs: $N=10$.}

The spin-orbit coupling term, $V_{\text{SOC}}$, captures the influence of the DNA helical geometry
\begin{equation}
V_{\text{SOC}} = \sum_{\beta j \sigma \sigma’} i \gamma_{\beta} \Lambda_{\sigma \sigma’}^{\beta j} d_{\beta j \sigma}^{\dagger} d_{\beta, j+1, \sigma’} + \text{h.c.},
\end{equation}
where  $\Lambda_{\sigma \sigma’}^{\beta j}$ are geometry-dependent matrices \cite{davis2024}.

The nuclear dynamics are described by a single classical mechanical coordinate $(p,x)$ with Hamiltonian
\begin{equation}
H_{\text{mech}}(p,x) = \frac{p^2}{2m} + \frac{1}{2} m \omega_0^2 x^2,
\label{Hmech}
\end{equation}
where $m = 5.9244 \times 10^5$ a.u. approximates the mass of a nucleotide, and $\omega_0 = 5$ meV represents the vibrational frequency of DNA's collective mechanical motion \cite{davis2024}.

The coupling between the electronic states and mechanical coordinate is assumed to be linear in terms of mechanical displacement
\begin{multline}
H_{\text{e-mech}}(x) = -\Big[\chi   \sum_{\beta j \sigma} d_{\beta j \sigma}^{\dagger} d_{\beta j \sigma} \\
	+\chi_1   \sum_{\beta j \sigma} \left( d_{\beta j \sigma}^{\dagger} d_{\beta, j+1, \sigma} + \text{h.c.} \right) \\
+ \chi_1  \sum_{j \sigma} \left( d_{A j \sigma}^{\dagger} d_{B j \sigma} + \text{h.c.} \right) \Big] x.
\label{emech}
\end{multline}
The first term describes the modulation of on-site energies, the second term accounts for the effect of mechanical motion on intra-strand hopping, and the third term captures inter-strand coupling modulations due to conformational changes. We assume coupling constants satisfy $\chi_1 = 0.2 \chi$. The value of $\chi$ is treated as a tuneable parameter to explore the impact of electron-mechanical motion coupling strength.

The coupling between DNA and electrodes is represented by
\begin{equation}
H_{ML} = \sum_{k \beta \sigma} \left( t_{L k, \beta 1} d_{L k \sigma}^{\dagger} d_{\beta 1 \sigma} + \text{h.c.} \right),
\end{equation}
\begin{equation}
H_{MR} = \sum_{k \beta \sigma} \left( t_{R k, \beta N} d_{R k \sigma}^{\dagger} d_{\beta N \sigma} + \text{h.c.} \right),
\end{equation}
where $t_{\alpha k, \beta 1}$ and $t_{\alpha k, \beta N}$ denote the tunnelling amplitudes between electrode single-particle states $\alpha k $ and the terminal nucleotides ($j=1$ or $j=N$) of strand $\beta$. It is assumed that both DNA strands are chemically bonded to the electrodes at their respective termini.

\subsection{Green's functions and self-energies}

The retarded component of leads self-energy is computed using wide-band approximation \cite{molecular_electronics_book}
\begin{equation}
\Sigma_{\alpha; \beta i \sigma, \beta' i' \sigma'}^R=  -\frac{i}{2} \Gamma_{\alpha; \beta i \sigma, \beta' i' \sigma'}.
\end{equation}
The imaginary part of the self-energy is given by
\begin{equation}
\Gamma_{\alpha; \beta i \sigma, \beta' i' \sigma'}= 2 \pi  \delta_{\beta \beta'} \delta_{i i'}\delta_{\sigma \sigma'} \sum_{k} t^{*}_{\alpha k, \beta i \sigma} \delta(\omega-\epsilon_{k \alpha}) t_{\alpha k, \beta i \sigma},
\end{equation}
and assumed to be $\omega$-independent. 
{  Matrix 
$\Gamma_{\alpha;\beta i\sigma,\beta' i'\sigma'}$ is diagonal in  strand ($\beta$), spin ($\sigma$), and nucleotide ($i$) indices
and is nonzero only for terminal nucleotides ($i=1,N$) directly coupled to the electrodes. 
This represents local, spin-independent coupling between the electrode states and the outermost base pairs of each strand. }
We use $\Gamma_{\alpha; \beta 1 \sigma, \beta 1 \sigma} =\Gamma_{\alpha; \beta N \sigma, \beta N \sigma} =0.1$ eV in our calculations.

The lesser and greater self-energies are computed as \cite{molecular_electronics_book}
\begin{equation}
\mathbf \Sigma_{\alpha }^<(\omega)=  i f_{\alpha}(\omega) \mathbf \Gamma_{\alpha}
\end{equation}
and
\begin{equation}
 \mathbf \Sigma_{\alpha}^>(\omega)=  -i (1-f_{\alpha}(\omega))\mathbf \Gamma_{\alpha},
\end{equation}
where $f_{\alpha}(\omega)$ are Fermi-Dirac occupation numbers for lead $\alpha$.
Note that throughout the paper, we utilise boldface letters to represent matrices in DNA spin-orbital space within equations.

 All components of the  Green's functions depend instantaneously on the DNA's mechanical degree of freedom and are defined using standard adiabatic relations \cite{molecular_electronics_book}:
\begin{equation}
\mathbf G^R(x ,\omega) = (\omega \mathbf I - \mathbf h(x) - \mathbf \Sigma^R)^{-1},
\end{equation}
\begin{equation}
\mathbf G^A(x ,\omega)= \big( \mathbf G^R(x ,\omega) \big)^\dag,
\end{equation}
\begin{equation}
\mathbf G^<(x ,\omega) = \mathbf G^R(x ,\omega)  \mathbf \Sigma^<(\omega) \mathbf G^A(x ,\omega)
\end{equation}
and
\begin{equation}
\mathbf G^>(x ,\omega) = \mathbf G^R( x ,\omega) \mathbf  \Sigma^>(\omega) \mathbf G^A(x ,\omega).
\end{equation}
In this section, we introduce matrix $\mathbf h(x)$, which represents the single-particle DNA Hamiltonian and includes the electron-mechanical motion term (\ref{emech}).

\subsection{Keldysh-Langevin molecular dynamics}
The mechanical degree of freedom of DNA is considered a classical variable within our approach. 
{  Quantum vibrational effects play an important role in inelastic electron transport through molecules \cite{molecular_electronics_book}.
In this work, however, the collective coordinate represents a slow, large-scale conformational motion of the DNA molecule , whose characteristic frequency (5~meV) is much smaller than   room temperature. 
In this limit, the motion is well described classically.  }
If we additionally assume that DNA’s conformational dynamics are slow relative to the tunnelling electrons, we can obtain a Langevin equation of motion \cite{pistolesi08,pistolesi10,Bode11, bode12,kosov12,todorov12,preston2020,kershaw2020}:

\begin{equation}
\frac{dp}{dt} = - \partial_x H_{\text{mech}} + F_{\text{e,neq}} - (\xi_e(x) + \xi_{\text{env}}) \dot x + \delta f(x,t).
\end{equation}

The Langevin equation above comprises a classical force, a non-equilibrium electronic force $F_{e,\text{neq}}$, a frictional force and its electronic viscosity $\xi_\text{e}(x)$ and the viscosity of the DNA environment $\xi_{\text{env}}$, and a stochastic force $\delta f(x,t)$.

 The electronic  force $F_e$ has the form
\begin{equation}
F_{\text{e}}( x ) = i \int \frac{d \omega}{2 \pi} \text{Tr} \Big[\partial_x \mathbf h \;  \mathbf G^<(   x, \omega) \Big],
\end{equation}
with the matrix $\partial_x \mathbf h$ is  the derivative of single-particle molecular Hamiltonian matrix.
The non-equilibrium electronic force which enters the Langevin equation is defined as
\begin{equation}
F_{\text{e,neq}}=F_{\text{e}}(x)-F_{\text{e,eq}}(x),
\label{fe}
\end{equation}
where it is assumed that the equilibrium potential energy surface is completely represented by  $ H_{\text{mech}}$ and as such the force computed using $- \partial_x H_{\text{mech}}$  must also have the equilibrium component (zero voltage bias) $F_{\text{e,eq}}(x)$ removed.

The electronic viscosity  depends on DNA's geometry and is given \cite{bode12,kosov12} by
\begin{multline}
\xi_{\text{e}}(   x)= \int \frac{d\omega}{2 \pi} \text{Tr} \Big[ \mathbf G^<(   x, \omega) \partial_x \mathbf h \partial_\omega \mathbf G^R(   x, \omega) \partial_x \mathbf h
\\
- \mathbf G^<(   x, \omega) \partial_x \mathbf h \partial_\omega \mathbf G^A(   x, \omega) \partial_x \mathbf h 
\Big].
\end{multline}
The stochastic force $\delta f(t)$ is modelled as a Markovian Gaussian variable with zero mean
\begin{equation}
\langle \delta f(t) \rangle = 0
\end{equation}
and delta-function variance
\begin{equation}
\langle \delta f(t) \delta f (t^{\prime}) \rangle = (D_{\text{e}}(x) +D_{\text{env}}) \delta (t - t^{\prime}),
\label{RARA}
\end{equation}
which contain contribution from electrons $D_{\text{e}}(x)$ and DNA environment $D_{\text{env}}$.
The electronic diffusion coefficient  can be expressed in terms of lesser and greater Green's functions \cite{bode12,kosov12}:
\begin{equation}
D_{\text{e}}(   x) = \int \frac{d \omega}{2 \pi} \text{Tr} \Big\{ \partial_x \mathbf h \mathbf G^{<}(   x, \omega) \partial_x \mathbf h \mathbf G^{>}(   x, \omega) \Big\}.
\end{equation}
We assume that environmental viscosity and diffusion coefficients are related via fluctuation-dissipation relation 
\begin{equation}
T= \frac{D_{\text{env}}}{2 \xi_{\text{env}}},
\end{equation}
where $T$ is the ambient temperature of the DNA junction. The ambient temperature is assumed to be identical to the temperature of electrons in the lead, and it is taken to be $T=300$ K in our calculations. 
We use the same environmental viscosity as in \cite{davis2024}:  $\xi_{\text{env}} = 2.228 \times 10^{-4}$ a.u. 

In the absence of position-dependent viscosities and random forces, the standard symplectic Langevin algorithm BAOAB \cite{leimkuhler2013}  can be employed to integrate the equations of motion. This algorithm involves inserting an exact solution to the Ornstein-Uhlenbeck process (OU) within a velocity Verlet algorithm. It is one of several algorithms based on splitting field updates.
In the presence of position-dependent dissipation and noise, the lack of fluctuation-dissipation necessitates a modification of the OU process. Instead of substituting the OU operator, a modified algorithm from Ref. \cite{sachs2017} is employed. This modified algorithm utilises a multistep Euler integration over the position-dependent fluctuation-dissipation to represent the OU operator.
The primary rationale for employing Euler’s method is to avoid a more intricate evaluation of the momentum update, thereby minimising computational time.

\begin{figure}[t!]

\includegraphics[scale=0.9]{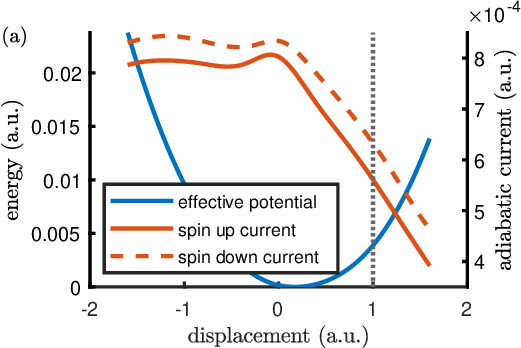}

\includegraphics[scale=0.9]{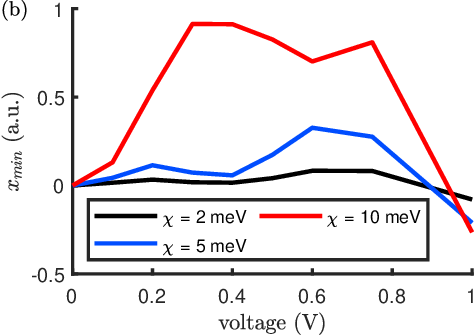}

\includegraphics[scale=0.9]{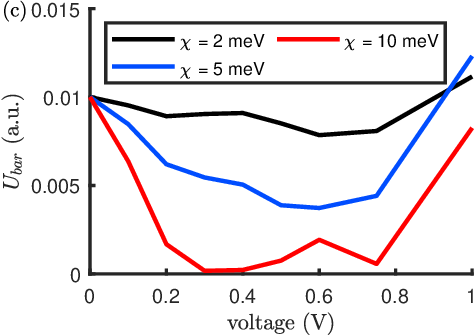}

\caption{
(a) Effective potential energy surface and spin-resolved currents as a function of displacement, with the critical instability threshold at $x =$ 1 a.u. (b) Voltage-dependent shift of the potential minimum, which moves toward destabilising displacements at stronger electron-mechanical motion coupling. (c) Corresponding barrier height at the critical displacement.}
\label{fig:Potentials}
\end{figure}

\section{Results}

\subsection{Electric current reshapes effective potential energy surface for DNA's mechanical motion}

We define the effective potential energy experienced by the DNA's mechanical degree of freedom as 
\begin{equation}
U(x)=\frac{1}{2}m\omega_{0}^{2}x^{2}-\int_{x_{0}}^{x}dxF_{\text{e,neq}}(x),\label{U}
\end{equation}
where the electronic force, denoted as $F_{\text{e,neq}}$, is given by (\ref{fe}) and the choice of $x_{0}$ is arbitrary.
The effective potential energy contains  the equilibrium part, which is the classical potential from (\ref{Hmech}),  and a modification to the potential due to current-induced force.
{ 
The effective potential $U(x)$ represents only the conservative component of the electronic force. 
It provides intuitive insight into how the electronic current reshapes the deterministic part of the energy landscape. 
However,   it is critical to include non-conservative viscous and stochastic forces, which are also explicitly computed in our Keldysh-Langevin molecular dynamics.}

Fig. \ref{fig:Potentials} illustrates how electronic current reshapes the effective potential of the DNA junction. Fig. \ref{fig:Potentials}(a) shows the effective potential energy surface together with the spin-resolved currents as a function of displacement. 

Our subsequent analysis of DNA mechanical stability requires that we select  the threshold value for mechanical motion which marks the onset of DNA's structural instability - we assume once this critical value is reached, the DNA experiences structural instability.
The critical threshold ($x_b = 1$ a.u.) is selected. 
{  Molecular dynamics studies of base-pair deformability  in double stranded DNA 
consistently show that conformational displacements of order 1-5~a.u  of nucleotides from Watson-Crick structure correspond to the onset of DNA mechanical instability \cite{zacharias2020, lindahl2017,mak2016}. 
In our coarse-grained model, the coordinate $x$ represents a collective conformational mode (bending, twisting, or stretching of multiple base pairs). 
A displacement of 1~a.u. along this collective coordinate therefore corresponds to the lower bound of the physical range for destabilisation of individual base pairs seen in atomistic studies. 
}

This choice is somewhat arbitrary given the collective nature of the DNA's mechanical coordinate in our model but can be physically motivated by typical ranges of realistic base-pairing potentials in DNA. The threshold $x_b = 1$ a.u.  has a corresponding critical potential energy $U(x_b)$  which will be treated as the energy barrier $U_{\text{bar}}$ that the mechanical degree of freedom has to overcome to reach an unstable configuration.

Fig. \ref{fig:Potentials} shows that the non-equilibrium electronic force shifts the effective potential minimum position and  modifies the barrier height. Fig. \ref{fig:Potentials}(a) also shows
the spin resolved adiabatic electric current  computed using the standard NEGF expression
\begin{multline}
J_{\alpha\sigma}(x)=\intop_{-\infty}^{\infty}\frac{d\omega}{2\pi}\text{Tr}\Big[\mathbf{G}^{<} (   x, \omega)\mathbf{\Sigma}_{\alpha\sigma}^{A}(\omega)+\mathbf{G}^{R}(x,\omega) \mathbf{\Sigma}_{\alpha\sigma}^{<} (\omega)
\\
-\mathbf{\Sigma}_{\alpha\sigma}^{<}(\omega) \mathbf{G}^{A}(   x, \omega)-\mathbf{\Sigma}_{\alpha\sigma}^{R}( \omega) \mathbf{G}^{<}(  x, \omega)\Big],\label{j0}
\end{multline}
for each DNA conformation $x$. Here $\mathbf{\Sigma}_{\alpha\sigma}$  is the lead $\alpha $ self-energy projected to spin  $\sigma$. We observe that the spin-up and spin-down channels remain only weakly differentiated, indicating that while current-induced forces dominate the potential landscape, back influence from the mechanical motion to electronic spin selectivity is modest.

Fig. \ref{fig:Potentials}(b) and Fig. \ref{fig:Potentials}(c) quantify how voltage and electron-mechanical motion coupling reshape the effective potential. At low voltages, the position of the potential minimum shifts toward the critical displacement, with the strongest coupling ($\chi = 10$ meV) showing the largest excursions. This shift reflects the destabilising influence of current-induced forces, which bias the coordinate toward structural instability. Correspondingly, the barrier height at the critical displacement decreases sharply, in some cases vanishing entirely for $\chi = 1$ meV, indicating a regime of near-instantaneous escape.

However, beyond a certain bias (typically between 0.2-0.6 V, depending on coupling strength), the trend reverses. The potential minimum moves back away from the critical displacement, and the effective barrier re-emerges. This recovery indicates that the applied current not only destabilises but can also restore partial stability by reorganising the potential landscape at higher bias. The barrier height correspondingly increases after its initial collapse, producing a non-monotonic dependence of stability on voltage.

This counter-intuitive re-stabilisation under strong bias will be explored in detail in section \ref{subsec: lifetimes}, where the non-equilibrium lifetime of DNA junction is directly computed.

\subsection{DNA non-equilibrium lifetime as mean first passage time}
 \label{subsec: lifetimes}
 
To estimate the DNA non-equilibrium lifetime in an electronic junction environment, we utilise the generalised Langevin equation to compute the lifetime relative to the equilibrium zero-voltage lifetime, which serves as a reference point. This computation is performed through numerical integration, incorporating electronic non-equilibrium, friction, and random forces derived from NEGF theory, specifically the Keldysh-Langevin molecular dynamics approach.

We define first-passage time as the time it takes for the DNA's mechanical coordinate, evolving according to the generalised Langevin equation, to reach a predefined critical displacement  $x_b$ for the first time. Starting from an initial state sampled from the stationary distribution, the trajectory is propagated under the influence of deterministic forces (mechanical restoring force, non-equilibrium electronic force), viscosity (environmental and electronic),  and stochastic forces (environmental and electronic noise). The first-passage time $\tau$ is then recorded as the elapsed simulation time until the coordinate $x(t)$ crosses the critical threshold $x_b$ (taken here as $x_b = 1$ a.u.) for the first time:
\begin{equation}
    \tau = \inf \left\{ t > 0 \;|\; x(t) \geq x_b \right\}.
\end{equation}
By repeating this procedure for an ensemble of trajectories (200 - 300 trajectories per data point), the mean first-passage time (MFPT)  $\langle \tau \rangle $ is obtained, which serves as a quantitative measure of the non-equilibrium lifetime of the DNA junction under current flow.

\begin{figure}
\centering
\includegraphics[scale=0.9]{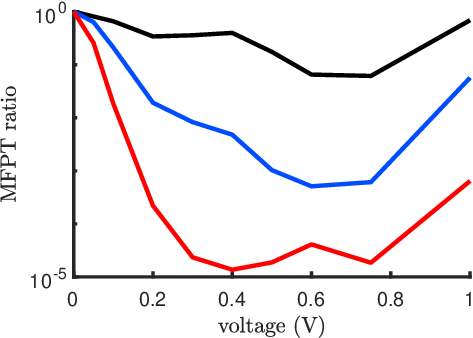}
\caption{Voltage dependence of the ratio of non-equilibrium MFPT to equilibrium MFPT for different  electron - mechanical motion  coupling strength: 
$\chi = 2$ meV (black), $\chi = 5$ meV (blue), $\chi = 10$ meV (red).}

\label{fig:FPT}
\end{figure}

Fig. \ref{fig:FPT} presents the voltage dependence of the ratio of non-equilibrium to equilibrium MFPTs  $\langle \tau(V) \rangle/ \langle \tau(V=0) \rangle$. At low voltages, all couplings show a rapid reduction in first-passage times, with the strongest coupling ($\chi = 10$ meV) producing non-equilibrium  lifetimes several orders of magnitude shorter than equilibrium. This behaviour reflects the destabilising influence of current-driven forces, which shift the potential minimum and reduce barrier heights (cf. Fig. \ref{fig:Potentials}).

However, a different feature emerges at higher voltages: instead of continuing to decrease, the MFPTs  begin to increase again. This recovery, most evident for $\chi = 10$ meV, is counter-intuitive given the higher kinetic energy injected into the system. The explanation lies in the reorganisation of the effective potential landscape - as voltage increases further, the position of the minimum moves back toward equilibrium and barrier height re-emerges, partially restoring stability (cf. Fig. \ref{fig:Potentials} ).

Thus, the non-monotonic voltage dependence of MFPTs highlights a subtle interplay between current-induced heating, potential energy surface reshaping, and spatially dependent dissipation: while moderate voltages destabilise DNA, higher voltages can paradoxically enhance stability by restoring effective energy barriers to prevent DNA denaturation.

\subsection{Landauer blowtorch effect emerges from localised electronic fluctuating and dissipating forces}

The Landauer blowtorch effect refers to the {\it kinetic}  destabilisation/stabilisation of a system when local fluctuations are spatially inhomogeneous. Even if the average potential landscape favours stability, regions of enhanced local noise can effectively lower barriers and promote escape, as if a ``blowtorch” were applied selectively along the reaction coordinate. The reverse is also observed at low voltages where regions of high temperature and moderate viscosity lead to diminished local density, which if located at the critical point can impede escapes and, thus, enhance stability.  Within the Langevin framework, this is captured by a coordinate-dependent effective temperature $T(x)$ defined through the generalised fluctuation-dissipation relation \cite{preston2021}
\begin{equation}
    T(x) = \frac{D_e(x) + D_{\text{env}}}{2(\xi_e(x) +\xi_{\text{env}})}.
    \label{Tx}
\end{equation}
{  Eq.(\ref{Tx}) provides a definition of the local effective temperature $T(x)$ rather than a true fluctuation-dissipation theorem. 
Only in equilibrium - that is, when the left and right chemical potentials are equal - does Eq.(\ref{Tx})  reduce to the genuine fluctuation-dissipation relation. 
In this limit, the $x$-dependences of the numerator and denominator cancel, and $T(x)$ coincides with the electrode temperature. }
The presence of non-equilibrium electronic viscosity and random forces result into inhomogeneous effective temperature $T(x)$.
When $T(x)$ varies significantly with $x$, the system no longer behaves as having a single temperature, leading to locally biased fluctuations and dissipation.

\begin{figure}

\includegraphics[scale=0.9]{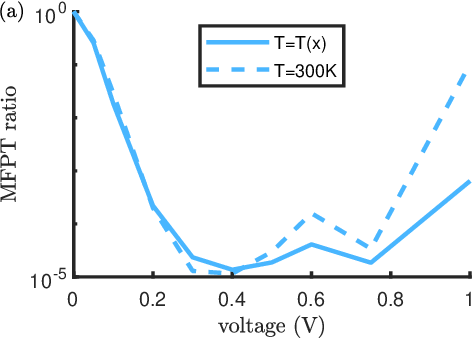}

\includegraphics[scale=0.9]{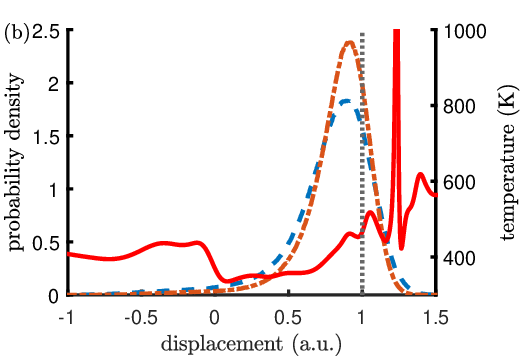}

\includegraphics[scale=0.9]{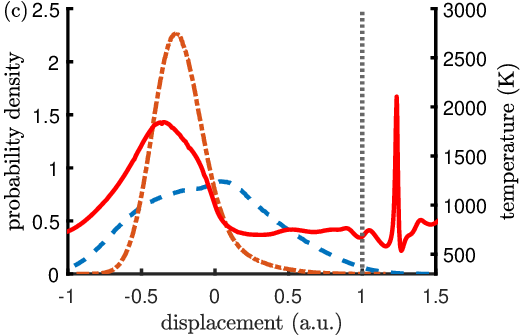}

\caption{
\
(a) Voltage dependence of MFPTs ratio computed with coordinate-dependent effective temperature $T(x)$ (solid) that means including electronic friction and fluctuating forces, compared to constant-temperature dynamics at $T=300$ K (dashed). 
(b) Probability densities from trajectories at $\chi = 10$ meV and $V = 0.3$ V. Including electronic viscosity and noise (blue dashed) reduces the probability near the threshold critical displacement compared to the distribution from non-equilibrium electronic forces and environmental dissipation alone (orange dashed). The locally elevated effective temperature $T(x)$ (red) illustrates the stabilising blowtorch effect.
(c) Same comparison at $V = 1$ V, showing even stronger broadening and access to high-displacement states as $T(x)$ rises sharply near the critical threshold displacement  -- destabilising blowtorch effect.
}
\label{fig:Blowtorch}
\end{figure}

Fig. \ref{fig:Blowtorch} highlights the role of coordinate-dependent dissipation and noise, manifesting as a Landauer-type blowtorch effect. 
Fig. \ref{fig:Blowtorch}(a) shows the consequences of spatially inhomogeneous effective temperature for dynamics. MFPTs are shorter for $V>0.5$ V  when electronic viscosity and noise are included (solid line), demonstrating that the blowtorch effect reduces stability beyond the influence of non-equilibrium electronic  forces alone. However, at the intermediate voltage range 0.2 -- 0.5 V the effect is opposite  and we observe extra stabilisation of the junction due to blowtorch effect.

Fig. \ref{fig:Blowtorch}(b) compares trajectory probability densities  and shows the spatial dependence of the effective temperature for $\chi = 10$ meV at $V = 0.3$ V. The steep increase in $T(x)$ as the coordinate approaches the critical displacement illustrates how locally enhanced noise intensity suppresses barrier crossing even when the average system temperature is increased by roughly 50\%. This spatial temperature gradient effectively increases the activation barrier, reinforcing the counter-intuitive observation that stability can emerge  from uneven distribution of heating along the reaction coordinate.

Fig. \ref{fig:Blowtorch}(c) compares trajectory probability densities with and without the inclusion of electronic viscosity and noise; the figure also shows the spatial dependence of the effective temperature for $\chi = 10$ meV at $V = 1$ V. When only the electron force and environmental fluctuation–dissipation are included (orange dashed), the distribution is narrowly localised around the potential minimum. By contrast, the full treatment including electronic viscosity and noise (blue dashed) broadens the distribution, enabling access to higher-energy displacements even in the presence of a confining potential. This is a direct signature of the blowtorch effect: non-uniform dissipation and stochastic forces effectively create spatial variations in “temperature”, making escape pathways more accessible.

\subsection{Apparent temperature of DNA's non-equilibrium mechanical motion and average electronic current}

\begin{figure}

\includegraphics[scale=0.9]{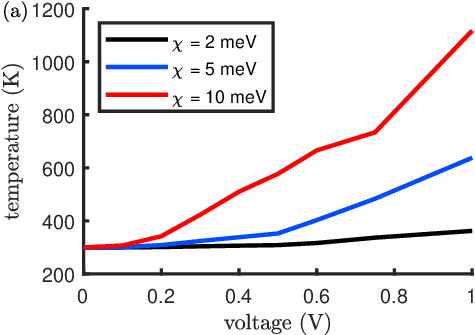}

\includegraphics[scale=0.9]{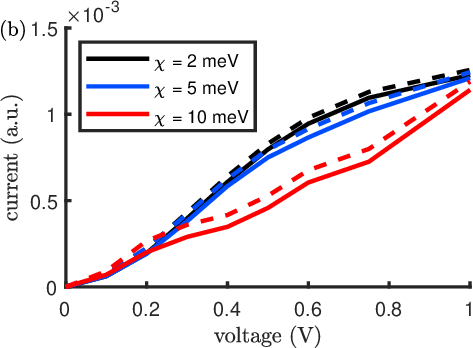}

\caption{(a) Trajectory-averaged apparent temperature of the DNA mechanical coordinate (computed as average kinetic energy) as a function of applied voltage for different electron-mechanical motion couplings.
 (b) Corresponding trajectory-averaged current through the DNA junction. Spin polarisation  $<14\%$  is evident in the difference between spin up (solid) and spin down (dashed). }
\label{fig:Temperature and current}
\end{figure}

Trajectory-averaged quantities are computed over a single 20 ns trajectory where configurations with $x$ greater than the structural instability threshold $x_b$  are not included in the statistics. The inclusion of electronic contributions to the fluctuating force and viscosity substantially elevates the apparent temperature  of the DNA’s mechanical coordinate which would be 300 K otherwise. 
As shown in Fig. \ref{fig:Temperature and current}(a), the apparent temperature rises sharply with applied bias, reaching more than four times the ambient value for the strongest electron-mechanical motion coupling ($\chi$ = 10 meV). Weaker couplings ($\chi$ = 2 meV, $\chi$ = 5 meV) display slower growth, illustrating the non-linear dependence of heating on coupling strength. This demonstrates that tunneling electrons act as an efficient energy pump into DNA mechanical motion, with even modest increases in $\chi$ leading to dramatic enhancement of DNA's mechanical kinetic energy under non-equilibrium conditions.

Fig. \ref{fig:Temperature and current}(b) presents the corresponding trajectory-averaged current.
 In all cases, the current increases with applied voltage before saturating near 1 V. The saturation magnitude is comparable across couplings, but the low-bias conductance decreases with increasing $\chi$, indicating that stronger coupling of electronic degrees of freedom to DNA's mechanical motion suppresses coherent quantum transport. Importantly, despite similar saturation currents, higher $\chi$ values generate significantly greater heating, showing that enhanced energy transfer efficiency, rather than current magnitude alone, drives the elevated effective temperatures. Together, these results establish that current-induced heating and suppressed transport are two interconnected signatures of strong coupling between quantum electrons and classical structural motion  in DNA junctions.
The spin  polarisation is $<14\%$ and does not exhibit substantial deviations from the previous work  where electronic friction and electronic  fluctuating force were not included into the Langevin  equation \cite{davis2024}.

Taken together, these results highlight the complex role of tunnelling electrons: they not only  reshape the effective potential through electronic force, as seen in Fig. \ref{fig:Potentials}, but also produce localised fluctuations and dissipation as seen in Fig. \ref{fig:Blowtorch}. Furthermore, they deposit a significant kinetic energy into the mechanical degree of freedom via non-equilibrium heating as depicted in Fig. \ref{fig:Temperature and current}.

\section{Conclusion}
We have applied non-equilibrium Keldysh-Langevin molecular dynamics to study the mechanical properties of  DNA in a molecular junction under electronic current. The simulations reveal that tunnelling electrons reshape the effective potential energy surface, modify fluctuation-dissipation balances, and generate spatially varying effective temperature (Landauer blowtorch effect). These effects manifest in three key findings.

First, the non-equilibrium lifetime of DNA is not a monotonic function of voltage: while moderate bias strongly destabilises the junction, higher voltages can paradoxically restore stability through reorganisation of the potential energy surface and energy barrier re-emergence. Second, the inclusion of electronic viscosity and fluctuating forces reveals an emergence of the Landauer blowtorch effect, where local effective temperatures can enhance or suppress barrier crossing depending on the applied voltage bias. Third, tunnelling electrons act as a powerful energy pump, raising the apparent  temperature of DNA far beyond the ambient conditions and suppressing coherent quantum transport.

Together, these results establish that the interplay of current-induced forces, dissipation, and noise governs the non-equilibrium lifetime of  DNA in nanoscale electronic junction environment. Beyond DNA, the approach and phenomena described here apply broadly to current-driven biomolecular and molecular electronic junctions, offering insights into how electronic currents can both destabilise and protect nanoscale structures.

\begin{acknowledgments}
 The authors thank Riley Preston for many valuable discussions.
 \end{acknowledgments}

\clearpage

\end{document}